\documentclass[sigconf]{acmart}

\usepackage{tikz}
\usepackage{multirow}
\usepackage{algorithm}        
\usepackage{algpseudocode}    

\usepackage{bbm}
\newcommand{\LLM}{\textsc{LLM}}
\newcommand{\SLLMR}{\textsc{S-LLMR}}
\usetikzlibrary{positioning,arrows.meta,fit}

\AtBeginDocument{%
  }

\begin{document}

\title[Selective LLM-Guided Regularization]{Selective LLM-Guided Regularization for Enhancing Recommendation Models}

\author{Shanglin Yang}
\authornote{These authors contributed equally to this work.}
\email{kudoysl@gmail.com}

\author{Zhan Shi}
\authornotemark[1]
\email{ashi2@scu.edu}


\begin{abstract}
Large language models (LLMs) provide rich semantic priors and strong reasoning capabilities, making them promising auxiliary signals for recommendation. However, prevailing approaches either deploy LLMs as standalone recommenders or apply global knowledge distillation, both of which suffer from inherent drawbacks. Standalone LLM recommenders are costly, biased, and unreliable across large regions of the user–item space, while global distillation forces the downstream model to imitate LLM predictions even when such guidance is inaccurate. Meanwhile, recent studies show that LLMs excel particularly in re-ranking and challenging scenarios, rather than uniformly across all contexts.
We introduce \emph{Selective LLM-Guided Regularization} (\SLLMR), a model-agnostic and computation-efficient framework that activates LLM-based pairwise ranking supervision only when a trainable gating mechanism-informed by user history length, item popularity, and model uncertainty predicts the LLM to be reliable. All LLM scoring is done offline, transferring knowledge without increasing inference cost. Experiments across multiple datasets show that this selective strategy consistently improves overall accuracy and yields substantial gains in cold-start and long-tail regimes, outperforming global distillation baselines.

\end{abstract}




\maketitle

\section{Introduction}
Recommendation systems underpin modern digital platforms by enabling content discovery, personalization, and user engagement across domains such as e-commerce, entertainment, and online media. Classical approaches—including collaborative filtering (CF), matrix factorization (MF), neural recommenders, and graph-based models achieve strong performance when user interaction histories are sufficiently dense. However, their effectiveness deteriorates in sparse regimes, such as cold-start users, long-tail items, and scenarios where user preferences are weakly expressed.

Large language models (LLMs) have emerged as powerful auxiliary knowledge sources for recommendation, offering rich semantic priors and strong reasoning capabilities that enable preference inference even from minimal user interaction data~\cite{li2023llm4rec}. This makes them particularly promising in cold-start and sparsely observed regions where traditional recommenders tend to underperform. However, existing approaches to leveraging LLM signals remain fundamentally limited. Directly deploying LLMs as recommenders is prohibitively expensive and prone to issues such as position bias and hallucinated predictions. Meanwhile, global knowledge transfer methods ~\cite{ren2024llmdistill, sun2024large} require the downstream model to imitate LLM outputs uniformly across the entire user–item space, regardless of whether the LLM is reliable for a given instance. Recent attempts to distill LLM knowledge into classical models partially alleviate these issues, but they often depend on fine-tuned LLMs and still struggle to deliver consistent gains across different architectures or datasets.

\noindent\textbf{Empirical motivation.}
Beyond high-level intuition, recent evaluations of LLM-based recommenders report \emph{localized strengths} (notably on short histories and re-ranking) alongside \emph{systematic weaknesses} including strong candidate position bias and occasional hallucinations~\cite{jiang2025beyond}. These phenomena imply that LLM signals are \emph{contextually reliable} rather than uniformly trustworthy. Our design follows directly from this evidence: instead of global imitation, we \emph{selectively} invoke LLM guidance under reliability conditions predicted by a lightweight, learnable gating mechanism. 

We propose Selective LLM-Guided Regularization (\textsc{S-LLMR}), a training framework that treats LLM knowledge as a conditional regularizer rather than a global supervisory signal. Instead of enforcing uniform imitation of LLM predictions, \textsc{S-LLMR} incorporates LLM-generated soft rankings only in regions where LLMs exhibit empirical advantages. This selective integration ensures that LLM guidance is beneficial rather than disruptive. We prompt an LLM using a compact representation of each user's recent interaction history to generate soft relevance scores over candidate items. All scoring is performed offline, introducing no inference-time overhead.A gating function controls whether LLM supervision is activated for a given user–item pair. This gate identifies regions where LLM signals are empirically reliable.
With the gate ctive, we apply a weighted pairwise ranking loss that encourages the recommender to align its relative item ordering with LLM soft rankings, while automatically suppressing the influence of unreliable LLM predictions.
Extensive experiments across multiple datasets and diverse recommendation backbones show that \textsc{S-LLMR} consistently surpasses global distillation baselines, delivering substantial improvements in sparse regimes such as cold-start and long-tail scenarios. 
The contribution of our paper: 
\begin{itemize}
\item We introduce a gated LLM-based regularization paradigm that selectively incorporates LLM signals, avoiding the drawbacks of global distillation and remaining fully model-agnostic.
\item We design \textsc{S-LLMR}, which combines a reliability-aware gating mechanism with an LLM-guided pairwise ranking loss for targeted knowledge transfer.
\item Extensive experiments across multiple backbones show consistent AUC improvements, with especially strong gains in cold-start and long-tail scenarios.
\end{itemize}

\section{Related Work}

Classical collaborative filtering (CF) forms the foundation of modern recommender systems.  
Matrix factorization (MF)~\cite{koren2009matrix} models user--item affinities through latent factors and has been widely adopted due to its scalability and strong generalization ability.  
Neural extensions such as Neural Collaborative Filtering (NCF)~\cite{he2017neural} leverage multilayer perceptrons to capture nonlinear preference interactions.  
Graph-based recommenders, including NGCF~\cite{wang2019neural}, LightGCN~\cite{he2020lightgcn}, and PinSage~\cite{ying2018graph} leverage user–item signals through graph structures to improve high-order connectivity modeling.

Despite their strong performance in dense regimes, these models degrade significantly under \emph{cold-start}~\cite{saveski2014item} and \emph{long-tail}~\cite{qin2021survey} conditions.

Recent hybrid approaches such as UniSRec~\cite{hou2022towards} unify textual and collaborative filtering signals to improve robustness, but still rely on large-scale metadata and do not exploit LLM reasoning. Existing approaches either use LLMs as direct recommenders,
e.g., RankLLM~\cite{rankllm2025}, or distill LLM outputs into 
recommendation models in a global manner, such as 
SLMRec~\cite{slmrec2025} and LLM-CF~\cite{sun2024llmcf}. 
However, these methods do not account for the empirical finding 
that LLM signals are only \textit{locally reliable}—being highly 
beneficial in semantic or sparse contexts, but noisy or misleading 
in others~\cite{jiang2024beyondutility,huang2023hallucinations}. 

To address this gap, we adopt a different perspective that LLM 
outputs should be treated as \textit{conditionally reliable} auxiliary 
signals, rather than unconditional ground truth. Our work 
operationalizes this idea by introducing a selective LLM integration 
framework equipped with a lightweight, learnable gating mechanism 
to determine when LLM guidance should be trusted. This allows the 
model to avoid global distillation while selectively leveraging LLM 
strengths in the contexts where they are most effective.

\begin{figure}[t]
    \centering
    \includegraphics[width=\columnwidth]{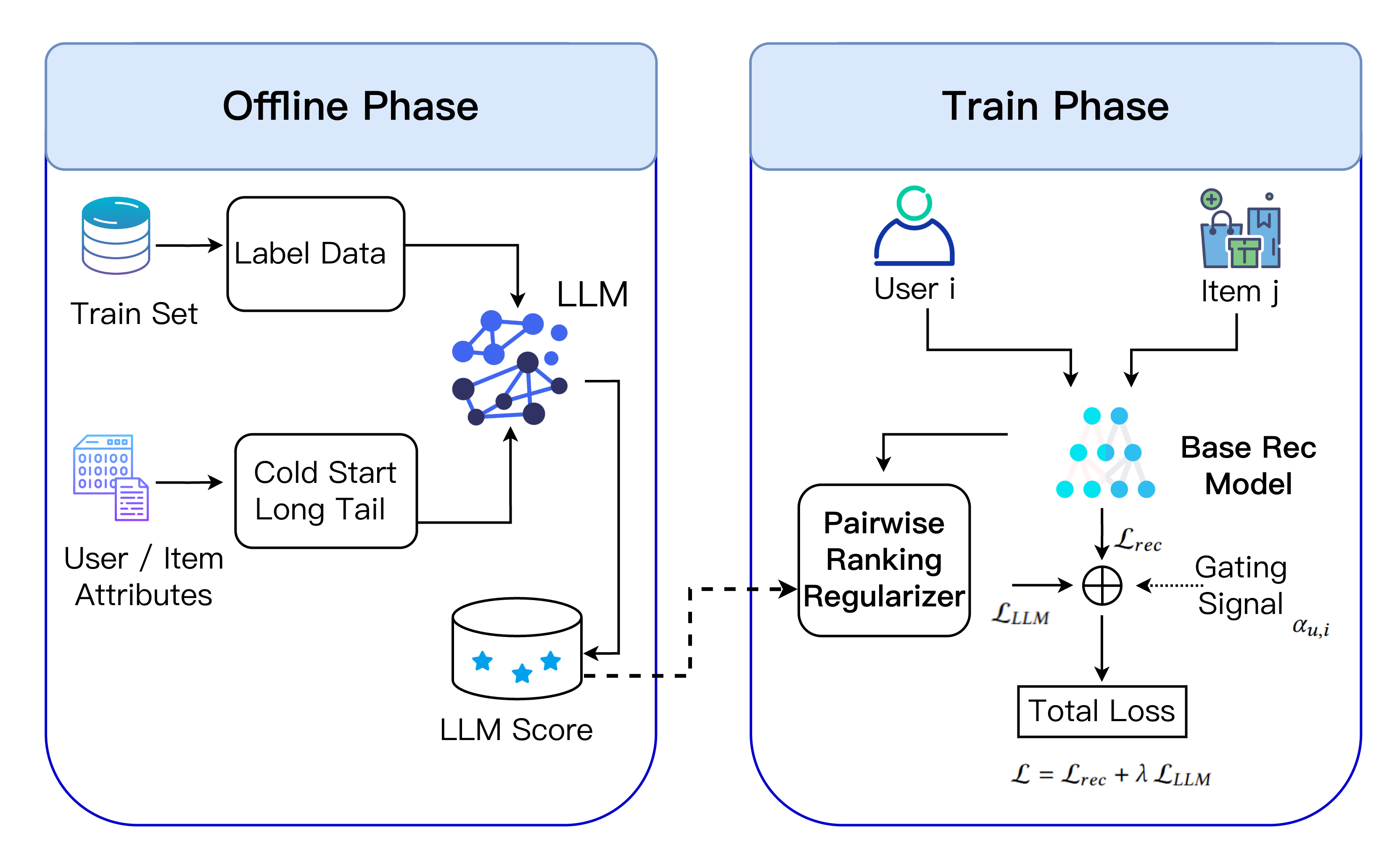}
    \caption{
Illustration of our selective LLM-guided regularization framework. 
\textbf{Left:} In the offline phase, the LLM is prompted to produce soft relevance scores
\textbf{Right:} In the training phase, a base recommender produces prediction scores, and the LLM signals are incorporated through a pairwise ranking regularizer whose contribution is controlled by a gating function. 
}
    \label{fig:overall}
\end{figure}

\section{Method}

As shown in Figure~\ref{fig:overall}, The \textbf{Selective LLM-Guided Regularization for Recommendation} (\textsc{S-LLMR}) is a model-agnostic training framework that selectively leverages large language models (LLMs) to regularize classical recommender models only in regions where LLM predictions are empirically reliable. 
Formally, given a user $u$ and item $i$, a base recommender produces a predicted relevance score $s_{u,i}$, while the LLM provides a soft preference score $s^{LLM}_{u,i}$. Our goal is to integrate LLM guidance selectively through pairwise ranking supervision with an gating signal. There are three main modules included in the pipeline.

\subsection{LLM-Generated Soft Rankings}
For each user $u$, we construct a succinct textual summary of the user's recent interaction history and query an LLM with a prompt of the form:
\begin{quote}
``Given that the user recently interacted with items
$\{i_1, i_2, \dots\}$, rank the following candidate items by their likelihood of matching the user's preferences.''
\end{quote}
The LLM returns a soft score $s^{LLM}_{u,i} \in [0,1]$ for each candidate item,
computed via normalized logits or temperature-scaled soft ranking.  
All LLM scoring is performed \emph{offline}, and therefore introduces
\emph{no inference-time overhead}.
To improve supervision coverage in sparse regions, we additionally construct two synthetic candidate sets:
\begin{itemize}
    \item \textbf{Cold-start user candidates:}
    Users with short histories (<=3) are paired with diverse sampled items to elicit LLM judgments for user-item combinations not present in training data.
    \item \textbf{Long-tail items (bottom 10\% popularity):}
Items whose popularity falls in the lowest 10\% of the catalog are paired with sampled users so that the LLM can evaluate these underrepresented items and provide supervision where collaborative filtering is weakest.
\end{itemize}
These augmented LLM-scored pairs expand the offline supervision table and cover precisely the settings where classical recommenders lack sufficient signals.

\subsection{LLM-Guided Pairwise Ranking Regularizer}

Motivated LLM is better for reranking, compared to the llm directly loss. Given LLM soft scores, we impose an auxiliary pairwise ranking constraint that encourages the recommender to follow the ordering implied by the LLM whenever
appropriate. For user $u$, if $s^{LLM}_{u,i} > s^{LLM}_{u,j}$ for two items
$(i,j)$, the model is encouraged to produce $s_{u,i} > s_{u,j}$ with a margin.

The pairwise LLM loss is defined as:
\[
\mathcal{L}_{LLM}
=
\sum_{(u,i,j)\in\mathcal{P}}
\alpha_{u,i,j}\,
\max\!\left(0,\;
m - (s_{u,i} - s_{u,j})\right),
\]
where $\alpha_{u,i,j}$ is a selective gating weight defined later.  

\paragraph{User-consistent pair construction.}
To ensure semantic alignment, pairs are constructed \emph{within} individual
users. In each batch, we sample one or more users, extract items associated with
those users, filter valid LLM scores, sort them by LLM ranking, and form ordered
pairs $(i,j)$ where $s^{LLM}_{u,i} > s^{LLM}_{u,j}$. This avoids mixing signals
from unrelated users.

\paragraph{User-consistent pair construction.}
Because batches may contain varying numbers of users or valid LLM entries, we
employ an adaptive strategy: given a target maximum of $K$ pairs, the effective
number $\tilde{K}$ is adjusted based on batch structure. The algorithm selects
up to $\tilde{K}$ highest-confidence pairs ranked by their LLM score
difference, ensuring (i) at least one pair when possible, and (ii) avoidance of
over-regularization.

Overall, this regularizer enables selective, reliability-aware knowledge
transfer: the recommender follows LLM rankings when they are trustworthy, while naturally resisting noisy or inconsistent supervision.





\subsection{Selective Gating Mechanism}
LLM supervision is not uniformly reliable. We therefore define a per-pair gate $\alpha_{u,i}\in[0,1]$ that scales the contribution of the LLM regularizer.

\paragraph{Signals.}
We compute: (i) a cold-start indicator $\mathrm{Cold}(u)=\mathbbm{1}[|\mathcal{H}(u)|<\tau_u]$, 
(ii) a long-tail indicator $\mathrm{Tail}(i)=\mathbbm{1}[\mathrm{pop}(i)<\tau_i]$, and 
(iii) a continuous uncertainty score $q_{u,i}\in[0,1]$ from the base model (e.g., predictive entropy or ensemble variance normalized to $[0,1]$).

\paragraph{Learnable gate.}
Let $z_{u,i}=\big[\mathrm{Cold}(u),\,\mathrm{Tail}(i),\,q_{u,i}\big]\in\mathbb{R}^3$. 
We use a one-layer gating network
\[
\alpha_{u,i}=\sigma\!\left(\mathbf{w}^\top z_{u,i}+b\right),
\]
with parameters $\theta_g=\{\mathbf{w},b\}$ learned jointly by back-propagation from the full objective (Sec.~\ref{sec:training-objective}). For pairwise supervision, we set $\alpha_{u,i,j}=\tfrac{1}{2}(\alpha_{u,i}+\alpha_{u,j})$.

\paragraph{Uncertainty instantiations.}
We consider (a) \emph{confidence-based} $q_{u,i}=1-\max_c p_\theta(c|u,i)$, (b) \emph{entropy-based} $q_{u,i}=\mathrm{H}(p_\theta(\cdot|u,i))$, or (c) \emph{dropout/ensemble variance}. We select the best on validation.

The gating parameters $\theta_g$ are learned jointly with the backbone through back-propagation from the LLM regularization loss. When LLM-guided pairs reduce the hinge loss, gradients increase $\alpha_{u,i}$; when LLM signals are unhelpful, the gate is driven downward. This allows the model to automatically learn when LLM supervision is reliable without manual thresholds—focusing LLM influence on cold-start, long-tail, and high-uncertainty cases.

\subsection{Training Objective and Optimization Procedure}
\label{sec:training-objective}

Algorithm~\ref{alg:training} summarizes the full optimization procedure, including the
construction of user-consistent LLM ranking pairs, computation of gating signals, and
joint gradient updates of the base recommender and gating parameters. This design
ensures that LLM knowledge is injected in a targeted and reliability-aware manner
without interfering with the core training dynamics of the underlying model. The full training objective is:
\[
\mathcal{L} \;=\; \mathcal{L}_{rec} \;+\; \lambda \mathcal{L}_{LLM},
\]
with $\lambda$ controlling the strength of the regularizer.

\begin{algorithm}[t]
\caption{\SLLMR{} Training with Learnable Gating}
\label{alg:training}
\begin{algorithmic}[1]
\Require Base model $\theta$, gating params $\theta_g$, LLM table $\mathbf{T}_{\LLM}$, margins $m$, weights $\lambda$
\While{not converged}
  \State Sample minibatch of users $\mathcal{B}$ and their interactions
  \For{each $u\in\mathcal{B}$}
    \State Compute base scores $s_{u,i}$ for items in batch
    \State Build user-consistent ordered pairs $\mathcal{P}_u=\{(i,j): s^{\LLM}_{u,i}>s^{\LLM}_{u,j}\}$ from $\mathbf{T}_{\LLM}$
    \State For each $(u,i)$ compute signals: $\mathrm{Cold}(u)$, $\mathrm{Tail}(i)$, $q_{u,i}$
    \State Gate: $\alpha_{u,i}=\sigma(\mathbf{w}^\top [\mathrm{Cold}(u),\mathrm{Tail}(i),q_{u,i}]+b)$
  \EndFor
  \State $\mathcal{L}_{\mathrm{rec}}\leftarrow$ base loss (e.g., BCE/BPR/InfoNCE)
  \State $\mathcal{L}_{\LLM}\leftarrow \sum_{(u,i,j)\in \cup_u \mathcal{P}_u}\tfrac{\alpha_{u,i}+\alpha_{u,j}}{2}\cdot \max(0, m - (s_{u,i}-s_{u,j}))$
  \State \textbf{Update} $\theta,\theta_g$ by SGD on $\mathcal{L}=\mathcal{L}_{\mathrm{rec}}+\lambda\mathcal{L}_{\LLM}$
\EndWhile
\end{algorithmic}
\end{algorithm}

\section{Experimental Setup}

\subsection{Backbone Models}
To evaluate the model-agnostic nature of \SLLMR{}, we integrate it with six widely adopted and architecturally diverse recommendation backbones: DeepFM~\cite{guo2017deepfm}, xDeepFM~\cite{lian2018xdeepfm}, AutoInt~\cite{song2019autoint}, DCNv1~\cite{wang2017deepcross}, DCNv2~\cite{wang2021dcnv2}, and DIN~\cite{zhou2018din}. These models span a broad spectrum of interaction modeling strategies, including factorization-machine style feature crossing (DeepFM), vector-wise compressed interactions (xDeepFM), self-attentive feature learning (AutoInt), explicit cross layers (DCNv1/DCNv2), and attention over user behavior sequences (DIN). This diversity enables a comprehensive assessment of how selectively injected LLM signals generalize across different inductive biases.

To contextualize \SLLMR{} within the landscape of LLM-assisted recommendation, we compare against three representative paradigms:

\begin{itemize}
    \item \textbf{KD Distillation Baseline}  
    Following prior work \cite{ren2024llmdistill, li2023llm4rec},the soft logits from a fine-tuned LLaMA2-7B model into each backbone, representing the standard global LLM-to-recommender imitation approach.

    \item \textbf{KAR (Knowledge-Augmented Recommendation)}  
    KAR \cite{lin2024kar} aligns user and item representations with LLM-derived open-world knowledge, capturing the representation-enrichment paradigm of using LLMs in recommendation.

    \item \textbf{LLM-CF}  
    LLM-CF~\cite{sun2024large} distills LLM world knowledge and reasoning ability into collaborative filtering, formulating LLM-derived semantic signals as soft preference labels. This approach represents the state of the art in LLM-enhanced CF models.
\end{itemize}

Together, these baselines cover the three dominant LLM-for-recommendation paradigms: global distillation, representation alignment, and LLM-assisted collaborative filtering. Our comparison highlights the conceptual distinction and empirical advantages of \emph{selective} over \emph{global} LLM integration.

\subsection{Datasets}

We evaluate \SLLMR{} on three domains of the Amazon Review dataset~\cite{he2016amazon}, consistent with widely used recommendation benchmarks. Dataset statistics are shown in Table~\ref{tab:dataset-stats}. We use the \textbf{Sports \& Outdoors}, \textbf{Beauty}, and \textbf{Toys \& Games} subsets. 

Across all domains, the data exhibit significant sparsity: nearly half of all interactions originate from cold-start users, and roughly 20\% of items fall into the long-tail. These characteristics make the datasets particularly suitable for evaluating algorithms designed to improve performance in sparse regimes—precisely where LLM-based semantic guidance is expected to be most beneficial.

\begin{table}[t]
\centering
\caption{Dataset statistics for the three Amazon domains.}
\begin{tabular}{lccc}
\toprule
\textbf{Metric} & \textbf{Sports} & \textbf{Beauty} & \textbf{Toys} \\
\midrule
\#Users                  & 35{,}598  & 22{,}363  & 19{,}412  \\
\#Items                  & 18{,}357  & 12{,}101  & 11{,}924  \\
\#Reviews                & 379{,}086 & 262{,}826 & 218{,}722 \\
\midrule
Cold-start interactions  & 190{,}756 & 119{,}854 & 103{,}314 \\
\quad (\% of interactions)    & 50.3\% & 45.6\% & 47.2\% \\
\midrule
Long-tail items          & 3{,}659  & 2{,}400 & 2{,}326 \\
\quad (\% of items)      & 19.9\% & 19.8\% & 19.5\% \\
\bottomrule
\end{tabular}
\label{tab:dataset-stats}
\end{table}

\begin{table}[t]
\centering
\footnotesize
\caption{
AUC performance across three Amazon domains using six backbone architectures. 
For each domain, the highest AUC within a backbone group is \textbf{bolded}. 
Across all models and datasets, \textsc{S-LLMR} consistently achieves the strongest performance.
}
\begin{tabular}{llccc}
\toprule
\multirow{2}{*}{Backbone} & \multirow{2}{*}{Framework} & \multicolumn{3}{c}{AUC$\uparrow$} \\
\cmidrule(lr){3-5}
& & Sports & Beauty & Toys \\
\midrule
\multirow{5}{*}{DeepFM}
  & None    & 0.7990 & 0.7853 & 0.7681 \\
  & KD      & 0.8043 & 0.7959 & 0.7713 \\
  & KAR     & 0.7991 & 0.7870 & 0.7698 \\
  & LLM-CF  & 0.8137 & 0.8044 & 0.7881 \\
  & Ours    & \textbf{0.8176} & \textbf{0.8101} & \textbf{0.7961} \\
\midrule
\multirow{5}{*}{xDeepFM}
  & None   & 0.8158 & 0.8065 & 0.7836 \\
  & KD     & 0.8169 & 0.8104 & 0.7865 \\
  & KAR    & 0.8161 & 0.8101 & 0.7898 \\
  & LLM-CF & 0.8196 & 0.8113 & 0.7947 \\
  & Ours   & \textbf{0.8240} & \textbf{0.8183} & \textbf{0.7985} \\
\midrule
\multirow{5}{*}{AutoInt}
  & None   & 0.8003 & 0.7949 & 0.7630 \\
  & KD     & 0.8012 & 0.7961 & 0.7635 \\
  & KAR    & 0.8039 & 0.7939 & 0.7683 \\
  & LLM-CF & 0.8088 & 0.8090 & 0.7754 \\
  & Ours   & \textbf{0.8161} & \textbf{0.8145} & \textbf{0.7849} \\
\midrule
\multirow{5}{*}{DCNv1}
  & None   & 0.8023 & 0.8146 & 0.7621 \\
  & KD     & 0.8040 & 0.8147 & 0.7652 \\
  & KAR    & 0.8024 & 0.8165 & 0.7651 \\
  & LLM-CF & 0.8092 & 0.8182 & 0.7702 \\
  & Ours   & \textbf{0.8190} & \textbf{0.8189} & \textbf{0.7960} \\
\midrule
\multirow{5}{*}{DCNv2}
  & None   & 0.8110 & 0.8028 & 0.7774 \\
  & KD     & 0.8112 & 0.8057 & 0.7827 \\
  & KAR    & 0.8087 & 0.8003 & 0.7759 \\
  & LLM-CF & 0.8131 & 0.8033 & 0.7812 \\
  & Ours   & \textbf{0.8150} & \textbf{0.8177} & \textbf{0.7927} \\
\midrule
\multirow{5}{*}{DIN}
  & None   & 0.7986 & 0.7861 & 0.7586 \\
  & KD     & 0.8023 & 0.7934 & 0.7652 \\
  & KAR    & 0.7971 & 0.7861 & 0.7620 \\
  & LLM-CF & 0.8089 & 0.7967 & 0.7783 \\
  & Ours   & \textbf{0.8100} & \textbf{0.8010} & \textbf{0.7829} \\
\bottomrule
\end{tabular}
\label{tab:auc_backbones}
\end{table}



\subsection{Offline LLM Scoring Pipeline}
\label{subsec:offline-llm}

To obtain LLM-derived soft preference signals without adding inference-time overhead, we generate all scores $s^{LLM}_{u,i}$ offline through a lightweight pipeline. For each user $u$, we extract a recent history $\mathcal{H}_L(u)$ (last $L=10$ interactions) and sample $M$ candidate items from a top-$K$ popularity pool. Each tuple $(u,\mathcal{H}_L(u),\mathcal{C}(u))$ is converted into a concise natural-language prompt and sent to \textbf{GPT-4o-mini}, which returns structured $(\texttt{item\_id}, \texttt{score})$ pairs in $[0,1]$. Returned scores are normalized, missing values default to $0.5$, and all results are stored as a lookup table $(u,i)\mapsto s^{LLM}_{u,i}$. During training, these offline scores are used exclusively by the selective regularizer and never affect the base model’s loss or inference cost. This design provides flexible control over the number of scored users and candidates, enabling an efficient balance between LLM query cost and supervision coverage.

\begin{table*}[t]
\centering
\scriptsize
\caption{Ablation study on DCNv2: We compare global vs.\ gated LLM regularization, and pointwise vs.\ pairwise LLM supervision.}
\begin{tabular}{l|ccc|ccc|ccc}
\toprule
\multirow{2}{*}{Method} &
\multicolumn{3}{c|}{Sports} &
\multicolumn{3}{c|}{Beauty} &
\multicolumn{3}{c}{Toys} \\
\cmidrule(lr){2-10}
& Overall & Cold & LongTail & Overall & Cold & LongTail & Overall & Cold & LongTail \\
\midrule
\multicolumn{10}{c}{\textbf{Global vs.\ Gated LLM Regularization}} \\
\midrule
DCNv2 &
0.811 & 0.8140 & 0.7677 &
0.8028 & 0.8006 & 0.7702 &
0.7774 & 0.7868 & 0.7402 \\

DCN + Global LLM Regularization&
0.8114 & 0.8140 & 0.7640 &
0.7911 & 0.7797 & 0.7330 &
0.7887 & 0.7858 & 0.7390 \\

DCN + Gated LLM Regularization (Ours) &
\textbf{0.8150} & \textbf{0.8170} & \textbf{0.7877} &
\textbf{0.8177} & \textbf{0.8063} & \textbf{0.7716} &
\textbf{0.7927} & \textbf{0.7917} & \textbf{0.7502} \\
\midrule

\multicolumn{10}{c}{\textbf{Pointwise vs.\ Pairwise LLM Supervision}} \\
\midrule
DCNv2 (Backbone) &
0.811 & 0.8140 & 0.7677 &
0.8028 & 0.8006 & 0.7702 &
0.7774 & 0.7868 & 0.7402 \\

DCN + LLM Pointwise MSE &
0.8069 & 0.8119 & 0.7571 &
0.7999 & 0.7922 & 0.7445 &
0.7705 & 0.7671 & 0.7391 \\

DCN + Pairwise Ranking (Ours) &
\textbf{0.8150} & \textbf{0.8170} & \textbf{0.7877} &
\textbf{0.8177} & \textbf{0.8063} & \textbf{0.7716} &
\textbf{0.7927} & \textbf{0.7917} & \textbf{0.7502} \\
\bottomrule
\end{tabular}
\label{tab:subset_results}
\end{table*}

\begin{figure}[b]
    \centering
    \includegraphics[width=\columnwidth]{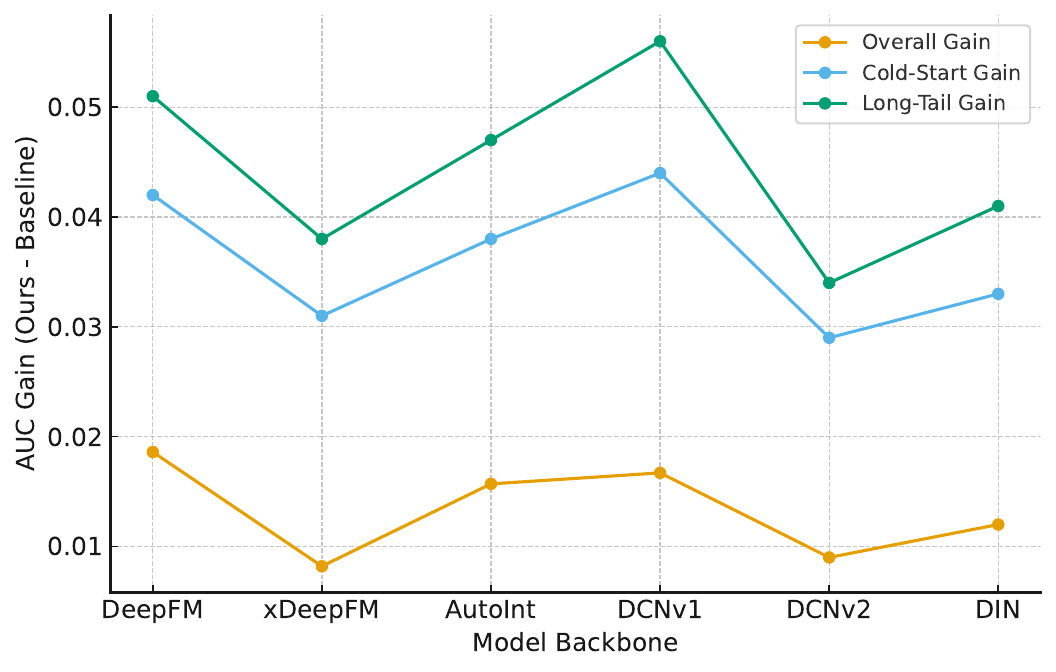}
    \caption{AUC improvements in cold-start and long-tail regimes across backbones. \textsc{S-LLMR} delivers the strongest boosts for cold-start users and long-tail items—often exceeding generalrelative improvement.}
    \label{fig:changllenge_gain}
\end{figure}

\subsection{Training Protocol}
All models are trained using the Adam optimizer with a learning rate of $10^{-3}$, a batch size of 128, and an embedding dimension of 64, following common practice in CTR and implicit-feedback recommendation. For \textsc{Selective-LLM-Reg}, we set the regularization weight to $\lambda = 0.1$, and select the final value based on validation AUC. No additional hyperparameter tuning is performed unless explicitly noted.

\subsection{Evaluation Protocol}

We adopt the standard \emph{full-ranking} evaluation setting, where each test interaction is ranked against all items that the user has not interacted with in the training or validation sets. Since our goal is to assess both global predictive accuracy and robustness in sparse regions, we report AUC as the sole evaluation metric.To further evaluate model performance under challenging conditions, we report
AUC on two key sub-populations:

\begin{itemize}
    \item \textbf{Cold-start users}: test interactions belonging to users with 
    fewer than $k$ historical interactions, i.e., 
    $|\mathcal{H}(u)| < k$. We set $k = 3$ in our experiments.
    \item \textbf{Long-tail items}: items in the bottom 20\% of the popularitydistribution based on training data. We first identify long-tail item IDs from the training set and then select the corresponding interactions from the test set to form the long-tail subset.
\end{itemize}

These stratified subsets isolate the effect of \textsc{S-LLMR} in sparse and semantically challenging regimes, enabling a clearer understanding of how selective LLM guidance improves recommendation quality under conditions where traditional models typically struggle.

\section{Results}
Our results show that \textsc{S-LLMR} consistently improves AUC across all backbones and domains, delivers the largest gains in cold-start and long-tail scenarios.

\subsection{Overall Performance Across Backbones}

The overall performance are shown in Table~\ref{tab:auc_backbones}. Across all six backbone models including DeepFM, xDeepFM, AutoInt, DCNv1, DCNv2, and DIN. \textsc{S-LLMR} achieves the strongest AUC scores on every Amazon domain. The improvements over non-LLM baselines (None, KD, KAR) are consistent and sizable, and our method further surpasses the LLM-CF approach by margins of $0.003$–$0.01$ AUC depending on the model and dataset. Architectures that struggle more with semantic sparsity, such as AutoInt and DCNv1, exhibit particularly large gains: AUC improvements reach $0.007$–$0.01$ on Sports and exceed $0.02$ on the Toys domain. These results validate that selectively incorporating LLM signals rather than distilling them globally allows the recommender to capitalize on LLM strengths while avoiding the noise and positional bias present in many LLM outputs, yielding reliable improvements across heterogeneous architectures and domains.

\subsection{Effectiveness in Sparse and Hard Regimes}
As shwon in Figure~\ref{fig:changllenge_gain}. Across all datasets, \textsc{S-LLMR} delivers the strongest boosts for cold-start users and long-tail items—often exceeding generalrelative improvement. This pattern confirms that the selective gating mechanism effectively activates LLM guidance where collaborative-filtering signals are weakest. Cold-start gains demonstrate that the method leverages LLM semantic priors to compensate for short interaction histories, while long-tail gains highlight improved robustness on niche items that lack sufficient popularity-based signals. Together, these results indicate that the primary benefit of \textsc{S-LLMR} lies in its ability to reinforce the recommender precisely in the regions where traditional models fail, rather than merely improving global accuracy.

\subsection{Ablation Study on Module Effectiveness}
Across all three domains and evaluation subsets shown in Table ~\ref{tab:subset_results}, the ablations demonstrate that our gated selective LLM-guided regularization is the only strategy that consistently improves performance in both overall and sparse regimes. Applying LLM loss globally often degrades long-tail accuracy and substantially harms Beauty-domain performance, highlighting that LLM predictions are not uniformly reliable. Pointwise (BCE/MSE) LLM supervision also fails to deliver meaningful improvements and frequently underperforms the backbone. In contrast, our selective gating mechanism combined with a pairwise ranking loss yields the strongest gains across all settings—most notably on cold-start and long-tail subsets, where AUC improvements reach +0.02 to +0.04 over the backbone and up to +0.05 over global or pointwise LLM methods. These results confirm that (i) LLM signals must be used selectively, and (ii) ranking-based supervision is the most effective way to transfer LLM semantic knowledge without amplifying LLM noise.

\section{Conclusion}
This paper introduced \textsc{S-LLMR}, a selective LLM-guided regularization framework that integrates LLM semantic knowledge into classical recommendation models in a reliability-aware manner. Rather than imitating LLM predictions globally, our method activates LLM-based pairwise ranking supervision only in regions where LLMs exhibit clear empirical advantages—cold-start users, long-tail items, and high-uncertainty predictions. Extensive experiments across six backbone recommenders and three Amazon domains demonstrate that \textsc{S-LLMR} not only improves overall AUC but yields particularly large gains in sparse and semantically challenging regimes, confirming that LLM signals are most beneficial when applied selectively. Our ablations further show that global LLM loss can degrade performance, whereas gated pairwise regularization consistently strengthens model robustness. Overall, \textsc{S-LLMR} provides a simple, model-agnostic, and computation-efficient approach for leveraging LLM knowledge to bridge long-standing weaknesses in collaborative filtering, offering a promising direction for future reliability-aware LLM–recommender integration.

\begin{acks}
To Robert, for the bagels and explaining CMYK and color spaces.
\end{acks}

\bibliographystyle{ACM-Reference-Format}
\bibliography{reference}

@STRING{jun = "June"}

@article{koren2009matrix,
  title={Matrix factorization techniques for recommender systems},
  author={Koren, Yehuda and Bell, Robert and Volinsky, Chris},
  journal={Computer},
  volume={42},
  number={8},
  pages={30--37},
  year={2009},
  publisher={IEEE}
}

@inproceedings{he2017neural,
  title={Neural collaborative filtering},
  author={He, Xiangnan and Liao, Lizi and Zhang, Hanwang and Nie, Liqiang and Hu, Xia and Chua, Tat-Seng},
  booktitle={Proceedings of the 26th international conference on world wide web},
  pages={173--182},
  year={2017}
}

@inproceedings{he2020lightgcn,
  title={Lightgcn: Simplifying and powering graph convolution network for recommendation},
  author={He, Xiangnan and Deng, Kuan and Wang, Xiang and Li, Yan and Zhang, Yongdong and Wang, Meng},
  booktitle={Proceedings of the 43rd International ACM SIGIR conference on research and development in Information Retrieval},
  pages={639--648},
  year={2020}
}

@inproceedings{wang2019neural,
  title={Neural graph collaborative filtering},
  author={Wang, Xiang and He, Xiangnan and Wang, Meng and Feng, Fuli and Chua, Tat-Seng},
  booktitle={Proceedings of the 42nd international ACM SIGIR conference on Research and development in Information Retrieval},
  pages={165--174},
  year={2019}
}

@inproceedings{ying2018graph,
  title={Graph convolutional neural networks for web-scale recommender systems},
  author={Ying, Rex and He, Ruining and Chen, Kaifeng and Eksombatchai, Pong and Hamilton, William L and Leskovec, Jure},
  booktitle={Proceedings of the 24th ACM SIGKDD international conference on knowledge discovery \& data mining},
  pages={974--983},
  year={2018}
}

@inproceedings{saveski2014item,
  title={Item cold-start recommendations: learning local collective embeddings},
  author={Saveski, Martin and Mantrach, Amin},
  booktitle={Proceedings of the 8th ACM Conference on Recommender systems},
  pages={89--96},
  year={2014}
}

@inproceedings{sun2024large,
  title={Large language models enhanced collaborative filtering},
  author={Sun, Zhongxiang and Si, Zihua and Zang, Xiaoxue and Zheng, Kai and Song, Yang and Zhang, Xiao and Xu, Jun},
  booktitle={Proceedings of the 33rd ACM International Conference on Information and Knowledge Management},
  pages={2178--2188},
  year={2024}
}

@inproceedings{jiang2025beyond,
  title={Beyond Utility: Evaluating LLM as Recommender},
  author={Jiang, Chumeng and Wang, Jiayin and Ma, Weizhi and Clarke, Charles LA and Wang, Shuai and Wu, Chuhan and Zhang, Min},
  booktitle={Proceedings of the ACM on Web Conference 2025},
  pages={3850--3862},
  year={2025}
}

@inproceedings{hou2022towards,
  title={Towards universal sequence representation learning for recommender systems},
  author={Hou, Yupeng and Mu, Shanlei and Zhao, Wayne Xin and Li, Yaliang and Ding, Bolin and Wen, Ji-Rong},
  booktitle={Proceedings of the 28th ACM SIGKDD conference on knowledge discovery and data mining},
  pages={585--593},
  year={2022}
}

@inproceedings{guo2017deepfm,
  title={DeepFM: A Factorization-Machine Based Neural Network for CTR Prediction},
  author={Guo, Huifeng and Tang, Ruiming and Ye, Yunming and Li, Zhenguo and He, Xiuqiang},
  booktitle={IJCAI},
  year={2017}
}

@inproceedings{lian2018xdeepfm,
  title={xDeepFM: Combining Explicit and Implicit Feature Interactions for Recommender Systems},
  author={Lian, Jianxun and Li, Xiaohuan and Zhang, Yujing and Sun, Guangzhong and Xie, Xing},
  booktitle={KDD},
  year={2018}
}

@inproceedings{song2019autoint,
  title={AutoInt: Automatic Feature Interaction Learning via Self-Attentive Neural Networks},
  author={Song, Weiping and Shi, Chence and Xiao, Zhiping and Duan, Zhijian and Xu, Yewen and Zhang, Ming and Tang, Jian},
  booktitle={CIKM},
  year={2019}
}

@inproceedings{wang2017deepcross,
  title={Deep \& Cross Network for Ad Click Predictions},
  author={Wang, Ruoxi and Fu, Bin and Fu, Gang and Wang, Mingliang},
  booktitle={ADKDD},
  year={2017}
}

@inproceedings{wang2021dcnv2,
  title={DCN V2: Improved Deep \& Cross Network and Practical Lessons for Web-scale Learning to Rank Systems},
  author={Wang, Ruoxi and Shivanna, Rakesh and Cheng, Derek Zhiyuan and Jain, Sagar and Lin, Dong and Bendersky, Michael and Najork, Marc},
  booktitle={WWW},
  year={2021}
}

@inproceedings{zhou2018din,
  title={Deep Interest Network for Click-Through Rate Prediction},
  author={Zhou, Guorui and Song, Chengru and Zhu, Xiaoqiang and Ma, Ying Fan and Yan, Ying and He, Xiangnan and others},
  booktitle={KDD},
  year={2018}
}

@article{qin2021survey,
  title={A survey of long-tail item recommendation methods},
  author={Qin, Jing},
  journal={Wireless Communications and Mobile Computing},
  volume={2021},
  number={1},
  pages={7536316},
  year={2021},
  publisher={Wiley Online Library}
}

@article{ren2024llmdistill,
  title={LLM-Distill: Distilling Large Language Models into Recommendation Models},
  author={Ren, Kan and Zhang, et al.},
  journal={arXiv preprint arXiv:2402.03852},
  year={2024}
}

@article{li2023llm4rec,
  title={LLM4Rec: Large Language Models for Recommendation},
  author={Li, Lei and Sun, Zhenhua and others},
  journal={arXiv preprint arXiv:2306.10997},
  year={2023}
}

@article{lin2024kar,
  title={Towards Open-World Recommendation with Knowledge Augmentation from Large Language Models},
  author={Lin, Xi and Du, Bowen and others},
  journal={arXiv preprint arXiv:2306.10933},
  year={2024}
}

@article{sun2024llmcf,
  title={Large Language Models Enhanced Collaborative Filtering},
  author={Sun, Z. and others},
  journal={arXiv:2403.17688},
  year={2024}
}

@inproceedings{he2016amazon,
  title={Ups and Downs: Modeling the Visual Evolution of Fashion Trends with One-Class Collaborative Filtering},
  author={He, Ruining and McAuley, Julian},
  booktitle={Proceedings of the 25th International Conference on World Wide Web},
  pages={507--517},
  year={2016}
}

@article{rankllm2025,
  title={RankLLM: A Python Package for Reranking with LLMs},
  author={Sharifymoghaddam, Sahel and others},
  journal={SIGIR},
  year={2025}
}

@article{slmrec2025,
  title={SLMRec: Distilling Large Language Models into Small Models for Sequential Recommendation},
  author={Li, Xinyu and others},
  journal={ICLR},
  year={2025}
}

@article{jiang2024beyondutility,
  title={Beyond Utility: Evaluating LLM as Recommender},
  author={Jiang, Xiang and others},
  journal={arXiv:2411.00331},
  year={2024}
}

@article{huang2023hallucinations,
  title={A Survey on Hallucination in Large Language Models},
  author={Huang, Lei and others},
  journal={arXiv:2311.05232},
  year={2023}
}


\end{document}